\documentclass[aps, pra, reprint, superscriptaddress]{revtex4-1}

\usepackage{amsmath}
\usepackage{newtxtext, newtxmath}

\usepackage[pdftex]{graphicx}
\usepackage{dcolumn}
\usepackage{bm}
\usepackage{soul}
\usepackage{hyperref}
\hypersetup{
    colorlinks=true,        
    linkcolor=blue,          
    citecolor=blue,        
    filecolor=magenta,      
    urlcolor=blue           
}
\usepackage{color}

\usepackage{braket}
\usepackage{xcolor}

\begin{document}

\title{Magnetoelastic couplings in the deformed Kagom\'{e} quantum spin lattice of volborthite}


\author{Akihiko~Ikeda}
\email[E-mail: ]{ikeda@issp.u-tokyo.ac.jp}
\affiliation{Institute for Solid State Physics, University of Tokyo, Kashiwa, Chiba, Japan}
\author{Shunsuke~Furukawa}
\email[E-mail: ]{furukawa@cat.phys.s.u-tokyo.ac.jp}
\affiliation{Department of Physics, University of Tokyo, Hongo, Tokyo, Japan}
\author{Oleg~Janson}
\affiliation{Institut f\"{u}r Festk\"{o}rperphysik, TU Wien, Wiedner Hauptstra$\beta$e 8-10, Vienna, Austria}
\affiliation{Institute for Theoretical Solid State Physics, IFW Dresden, Helmholtzstr. 20, 01069 Dresden, Germany}
\author{Yasuhiro~H.~Matsuda}
\author{Shojiro~Takeyama}
\affiliation{Institute for Solid State Physics, University of Tokyo, Kashiwa, Chiba, Japan}
\author{Takeshi~Yajima}
\author{Zenji~Hiroi}
\affiliation{Institute for Solid State Physics, University of Tokyo, Kashiwa, Chiba, Japan}
\author{Hajime~Ishikawa}
\affiliation{Institut f\"{u}r Funktionelle Materie und Quantentechnologien,  
Universit\"{a}t Stuttgart, Stuttgart, Germany}


\begin{abstract}
Microscopic spin interactions on a deformed Kagom\'{e} lattice of volborthite are investigated through magnetoelastic couplings.
A negative longitudinal magnetostriction $\Delta L<0$ in the $b$ axis is observed, which depends on the magnetization $M$ with a peculiar relation of $\Delta L/L \propto M^{1.3}$.
Based on the exchange striction model, it is argued that the negative magnetostriction originates from a pantograph-like lattice change of the Cu-O-Cu chain in the $b$ axis, and that the peculiar dependence arises from the local spin correlation.
This idea is supported by DFT+$U$ calculations simulating the lattice change and a finite-size calculation of the spin correlation, indicating that the recently proposed coupled-trimer model is a plausible one.
\end{abstract}

\maketitle

\begin{figure}[b!]
\includegraphics[scale=0.5]{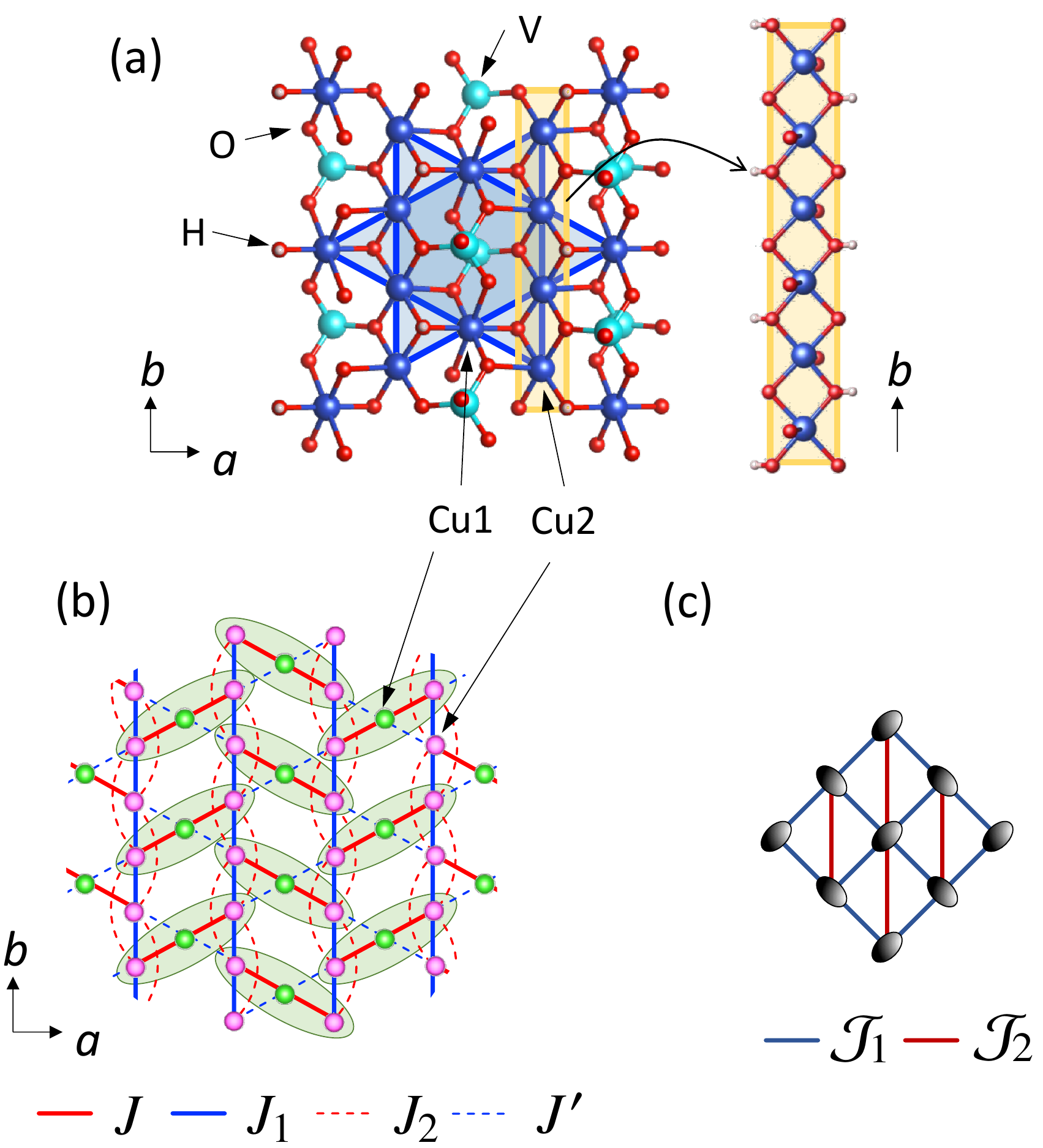}
\caption{
(a) Crystal structure of volborthite in the $a$-$b$ plane, in which Cu ions form a deformed Kagom\'{e} lattice (Left).
Magnification of  the Cu-O-Cu chain along the $b$ axis (Right).
Lattice parameters are based on Ref. \cite{IshikawaPRL}.
(b) Coupled-trimer model for volborthite \cite{JansonPRL}.
Blue and red bonds are ferromagnetic and antiferromagnetic couplings, respectively.
When $J\gg |J_{1}|$, $J_{2}$, $|J'|$, Cu2-Cu1-Cu2 can be regarded as a magnetic trimer with total spin $S_{\rm{tri}}=1/2$ up to the 1/3 magnetization plateau.
(c) Pseudospin-1/2 effective model, whose sites correspond to trimers in the original trimer model (see Ref. \cite{JansonPRL} for details).
Blue and red bonds are ferromagnetic and antiferromagnetic effective couplings, respectively.
\label{lattice}}
\end{figure}

The copper mineral volborthite Cu$_{3}$V$_{2}$O$_{7}$(OH)$_{2}\cdot2$H$_{2}$O is a fascinating example of a highly frustrated quantum magnet that exhibits a wealth of field-induced phenomena \cite{IshikawaPRL, YoshidaPRB}. 
In its magnetic layer, Cu ions possessing spin-1/2 moments form a deformed Kagome lattice as schematically shown in Fig. \ref{lattice}(a). 
Although this material was initially studied as a candidate for a Kagom\'{e} antiferromagnet with possible spin liquid behavior \cite{Hiroi, Bert}, 
it was later realized that the deformation of the lattice can lead to a significant spatial anisotropy in magnetic interactions \cite{YoshidaNC}. 
In particular, a microscopic spin model based on coupled trimers as shown in Fig. \ref{lattice}(b) has recently been proposed \cite{JansonPRL}, that  now attracts attention as it provides a mechanism for a field-induced spin nematic phase adjacent to the 1/3 magnetization plateau.
In the spin nematic phase, spin directors that break the in-plane rotational symmetry are formed as a result of the Bose-Einstein condensation of bi-magnon excitations \cite{Shannon, Hikihara}.

A key factor for emergence of the spin nematic phase is the competition between ferromagnetic and antiferromagnetic couplings in the effective model as schematically shown in Fig. \ref{lattice}(c).
In the coupled-trimer model, the strongest antiferromagnetic exchange coupling $J$ forms spin trimers with the low-lying $S_{\rm{tri}}=1/2$ and excited $S_{\rm{tri}}=3/2$ states.
In the magnetization process, all trimers are in the $S_{\rm{tri}}=1/2$ sector up to the 1/3 magnetization plateau, at which all timers are polarized with $S^{z}_{\rm{tri}}=1/2$.
Thus, the pseudospin-1/2 effective model up to the 1/3 magnetization plateau can be constructed, in which ferromagnetic $\mathcal{J}_{1}$ and antiferromagnetic $\mathcal{J}_{2}$ couplings compete on a triangular lattice as shown in Fig. \ref{lattice}(c).
Experimental determination of the respective exchange constants in Fig. \ref{lattice}(b) has however been elusive owing to the fact that most techniques observe magnetic moments rather than the respective interactions among them.

\begin{figure*}
\includegraphics[scale=0.42]{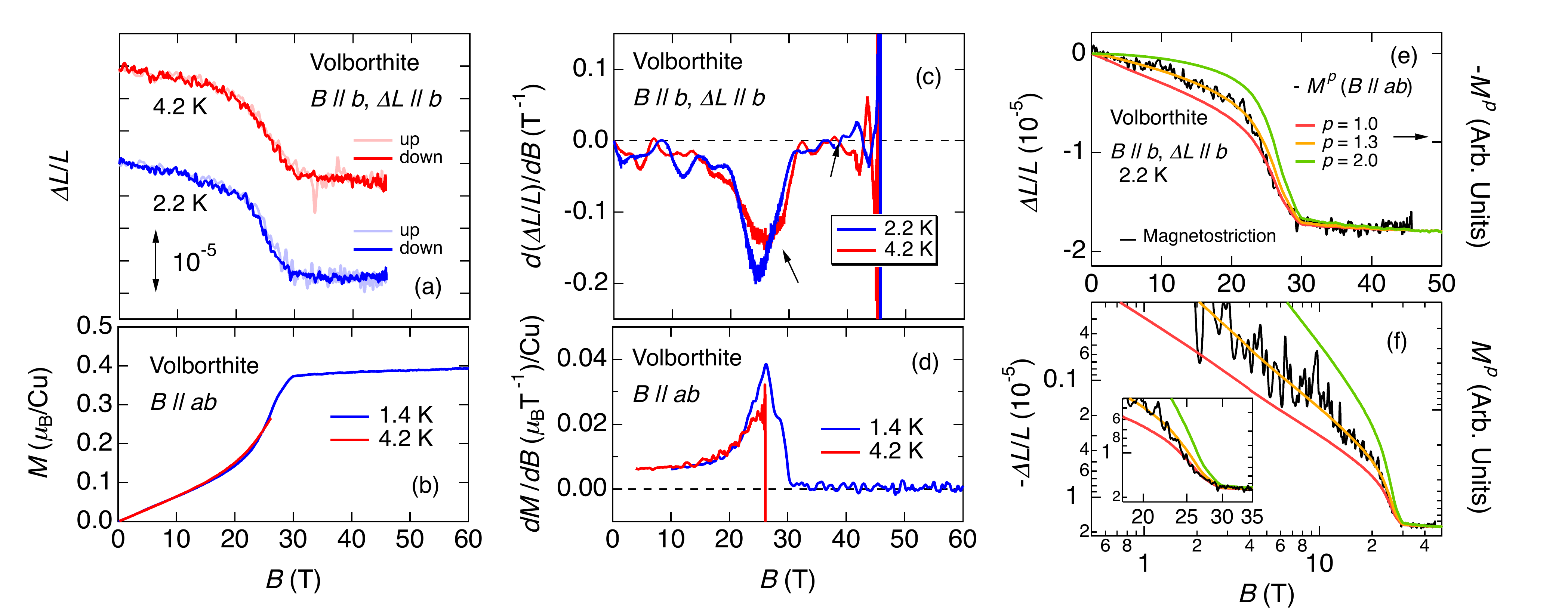}
\caption{ 
(a) Longitudinal magnetostriction of volborthite in the $b$ axis under 30 ms-pulsed high magnetic fields up to 45 T at 4.2 and 2.2 K.
(b) Magnetization curve at 4.2 and 1.4 K up to 27 and 60 T, respectively, in the $ab$ plane imported from Ref. \cite{IshikawaPRL}.
(c) Field derivative of the magnetostriction.
(d) Field derivative of the magnetization. 
(e) Magnetostriction at 2.2 K compared with magnetization $M$ at 1.4 K oriented in $B\parallel ab$, and $M^{p}$ with $p= 1.3$ and 2.0.
(f) Log-log plot of the same data. Inset is a magnification.\label{result}}
\end{figure*}

Magnetostriction is a unique measure of inter-site quanities in quantum spin systems such as the local spin correlations and the strain-dependences of the exchange couplings.
It thus provides ones with rare opportunities to access the information of exchange constants through the consideration of magnetoelastic couplings.
Furthermore, recent Faraday rotation measurements for volborthite have raised interesting possibilities of a structural transition and an associated change of a spin model at a high field \cite{Nakamura}, which implies the significance of magnetoelastic couplings.

In this paper, we report magnetostriction measurements of volborthite using a fiber Bragg grating (FBG) based strain measurement technique \cite{IkedaFBGHR}.
We observed small and negative magnetostriction in the crystallographic $b$ axis with a peculiar dependence of $\Delta L/L \propto M^{1.3}$.
They are rationalized with a pantograph-like lattice change of the Cu-O-Cu chain and the behavior of the local spin correlations with the aid of first principles calculations and an exact diagonalization study of the effective model in Fig. \ref{lattice}(c).

A selected single crystalline sample of volborthite \cite{IshikawaActa} with the dimension of $2 \times0.4 \times0.2$ mm$^{3}$ is glued to a FBG strain sensor along the $b$ axis.
Magnetostriction is measured using the FBG based strain measurement system, with a resolution of $\Delta L/ L \sim1\times10^{-6}$ where the optical filter method is employed as a detection scheme \cite{IkedaFBGHR}.
High magnetic fields are generated using a non-destructive pulsed magnet in IMGSL, ISSP, UTokyo, Japan.
All results reported here are for the longitudinal magnetostriction along the $b$ axis.

The results of the longitudinal magnetostriction measurements of volborthite in the $b$ axis at 4.2 and 2.2 K are shown in Fig. \ref{result}(a).
The magnetostriction is negative and the data for the up sweep and down sweep of the pulsed magnetic fields coincide with each other without hysteresis within the resolution, indicating that the magnetic phase transitions are continuous.
The magnitude of the magnetostriction shows a qualitatively similar behavior as the magnetization curve reported in Ref. \cite{IshikawaPRL} as shown in Fig. \ref{result}(b).  
In Fig. \ref{result}(c), the field derivatives of the magnetostriction, $d(\Delta L/L)/dB$, at 4.2 and 2.2 K are shown, where the temperature dependence is apparent.
Compared with the data at 4.2 K, the magnitude of the peak of $d(\Delta L/L)/dB$ becomes larger and the peak position shifts slightly to the lower field at 2.2 K.
This suggests that the plateau phase is more stable at 2.2 K, resulting in the lower entrance field to the plateau phase.
Similar enlargement of the peak is also observed in $dM/dB$ as shown in Fig. \ref{result}(d).
It should be noted that the peak of $dM/dB$ at 1.4 K has some shoulder structure, which is not seen in that of $d(\Delta L/L)/dB$ at 2.2 K.
This may be due to the possible spin nematic phase that appears below 2.2 K \cite{IshikawaPRL, YoshidaPRB}.
In Figs. \ref{result}(e) and \ref{result}(f), the data of $\Delta L/L$ is compared with the magnetization curves with the power of $p$.
It is apparently seen that the trend of $\Delta L/L$ data agrees best with $\Delta L/L \propto M^{p}$ with $p=1.3$, whereas $M^{p}$ curves with $p=1.0$ and $2.0$ clearly cannot be fitted to the data of $\Delta L/L$.

We discuss the origins of the negative magnetostriction and the relation of $\Delta L/L\propto M^{1.3}$ observed in volborthite in terms of  the exchange striction model.
As a possible origin of the negative magnetostriction, we propose a pantograph-like change of the Cu-O-Cu chain in the $b$ axis.
As for the dependence $\Delta L/L\propto M^{1.3}$, we argue that the local spin correlator is responsible.
Magnetoelastic couplings arising from on-site spin-orbit couplings, crystal field effects and Jahn-Teller effects are considered to be insignificant in the present system.
We note that, based on the coupled-trimer model \cite{JansonPRL}, it is reasonable to focus on the second strongest $J_{1}$ bond along the $b$ axis and neglect the magnetostriction in the strongest $J$ bond of in Fig. \ref{lattice}(b) up to the 1/3 magnetization plateau.
This is because the total spin of each trimer is fixed to the spin doublet $S_{\rm{tri}}=1/2$ up to the 1/3 magnetization plateau.
In the exchange striction model, magnetostriction is dependent on the spin correlation $\braket{\bm{S}_{i}\cdot \bm{S}_{j}}$ as discussed later, which is fixed at -1/2 in the above doublet sector.
Thus, one does not need to worry about the intra-trimer magnetostriction up to the 1/3 magnetization plateau.

\begin{figure}[h]
\includegraphics[scale=0.4]{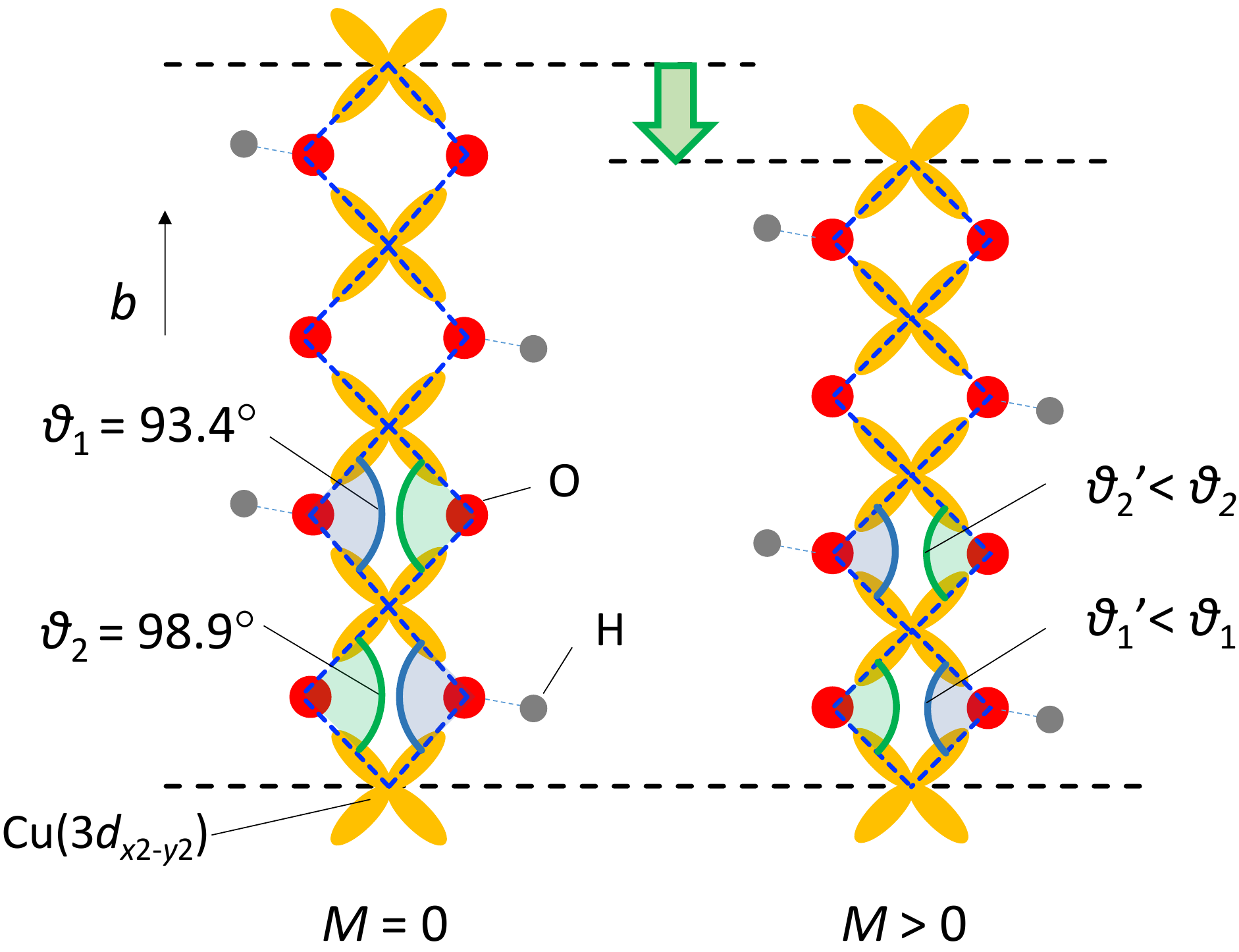}
\caption{Schematic drawing of a pantograph-like motion of the Cu-O-Cu chain in the $b$ direction of volborthite.
\label{model}}
\end{figure}

We consider the pantograph-like lattice change in the Cu chains in the $b$ axis to discuss the magnetoelastic coupling on the $J_1$ bonds.
As shown in Fig. \ref{model}, the 3$d_{x^{2}-y^{2}}$ orbital of each Cu2 site is connected to the 3$d_{x^{2}-y^{2}}$ orbital of the adjacent  Cu2 site through two paths of Cu-O-Cu bonds \cite{YoshidaNC}.  
On the left in Fig. \ref{model}, the lattice model at $M=0$ is shown, which is drawn from the lattice parameter at 55 K \cite{IshikawaPRL}, where the Cu-O-Cu bond angles are 93.4$^{\circ}$ and 98.9$^{\circ}$.
On the right, the lattice at $M>0$ is shown, where the Cu-O-Cu bond angles are reduced from the original values, approaching 90$^{\circ}$.
In Fig. \ref{model}, the way the crystal lattice changes mimics the motion of a pantograph, which was originally discussed for Cu dimers in the magnetostriction of SrCu(BO$_{2}$)$_{3}$ \cite{NarumiJPSJ, JaimePNAS, RadtkePNAS}.

The proposed model of the pantograph-like lattice modification is tested with density functional theory calculations with the on-site Coulomb term (DFT$+U$).
To calculate the evolution of the exchange integrals $J_{1}$, $J$, $J'$, and $J_{2}$ in Fig. \ref{lattice}(b), we vary the lattice parameter $b$ and calculate the DFT+$U$ total energies using the full-potential code FPLO \cite{Koepernik}.
For further details of the computational procedure, we refer the reader to Ref. \cite{JansonPRL}.
Note that the extremely low energy scale of the experimentally measured magnetostriction is beyond the reach of DFT total energy calculations.
To overcome this difficulty, we perform calculations for a largely enhanced lattice contraction.

\begin{figure}
\includegraphics[scale=0.52]{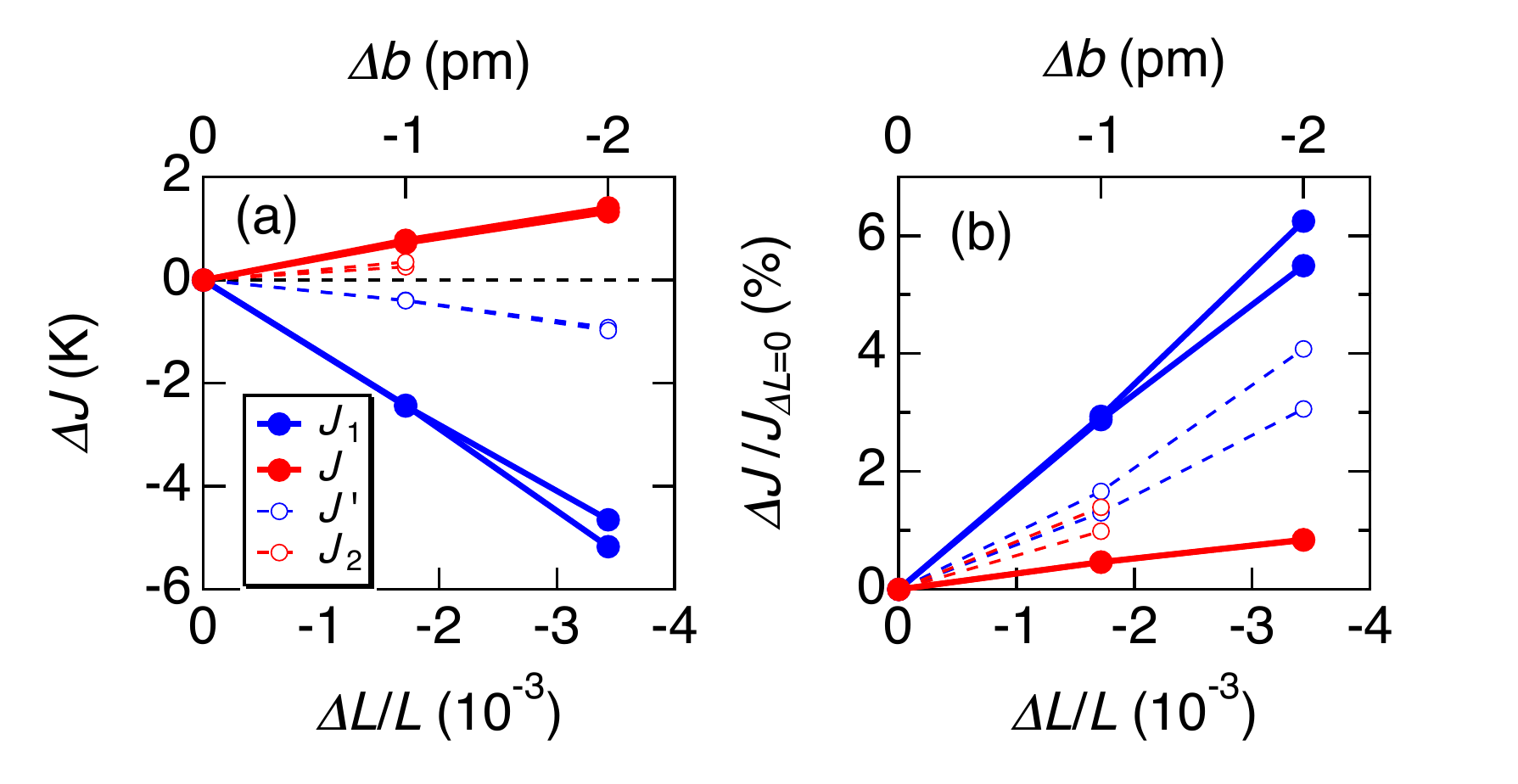}
\caption{(a) Changes and (b) relative changes in the exchange constants as a function of the decreasing lattice constant $b$ obtained with the DFT+$U$ calculations.
Blue- and red-colored symbols correspond to ferromagnetic (negative) and antiferromagnetic (positive) exchange interactions, respectively.
Two sets of data for each exchange constant are for the two structurally inequivalent layers.
\label{DFT}}
\end{figure}

The main outcome of the DFT+$U$ calculations is the enhancement of the ferromagnetic exchange constant $J_{1}$ upon a striction along the $b$ axis as shown in Fig. \ref{DFT}.
The lattice constant $b$ is the only parameter varied in the calculations.
The changes in the exchange constants $\Delta J$ and their ratios $\Delta J/J_{\Delta L=0}$ to the original values for different bonds are shown in Figs. \ref{DFT}(a) and \ref{DFT}(b).
The values of exchange constants at $\Delta b=0$ are, $J_{1}=-84.6$ $(-82.7)$ K, $J=156.3$ $(167.4)$ K, $J'=-30.0$ $(-24.0)$ K, $J_{2}=26.4$ $(25.1)$ K \cite{JansonPRL}, where the two values for each constant are for the two structurally inequivalent layers.
The initial value of $b$ is 5.8415 \AA \ \cite{YoshidaNC}.

The DFT$+U$ calculations qualitatively support our model of the pantograph-like lattice modification.
As can be seen in Fig. \ref{DFT}(b), both the ferromagnetic (blue-colored symbols) and antiferromagnetic (red-colored symbols) exchange constants are enhanced with decreasing $b$, which results from the enhanced overlap of the relevant orbitals responsible for the exchange bonds.
As seen in the absolute values [Fig. \ref{DFT}(a)] and the ratios of changes [Fig. \ref{DFT}(b)], enhancement of the exchange constant with decreasing $b$ is the most prominent for $J_{1}$ [Fig. \ref{lattice}(b)].
On the other hand, the changes in $J'$ and $J_2$ are relatively small, and the change in $J$ is expected to be irrelevant to the magnetostriction as discussed above.
These results support the idea that the change in $b$ originates primarily from the change in $J_{1}$ within the exchange striction model.

We, here, introduce a Hamiltonian based on the spin model in Fig. \ref{lattice}(b) as $\mathcal{H}_{\rm{es}}=J\sum_{i,j}\bm{S}_{i}\cdot\bm{S}_{j}+(J_{1}-p\epsilon)\sum_{i,j}\bm{S}_{i}\cdot\bm{S}_{j} + k\epsilon^{2}/2+J_{2}\sum_{i,j}\bm{S}_{i}\cdot\bm{S}_{j}+J'\sum_{i,j}\bm{S}_{i}\cdot\bm{S}_{j}+h_{z}\sum_{i}S^{z}_{i}$, where each summation is taken to satisfy the configuration of the exchange bonds in Fig. \ref{lattice}(b). 
We have introduced the elastic energy term with $\epsilon=\Delta L/L$ in the $b$ axis and a strain dependence to the $J_{1}$ bond.
This is a reasonable simplification considering the result of the DFT+$U$ calculation in Fig. \ref{DFT}(a).
By taking $dE/d\epsilon=0$ with $E=\braket{\mathcal{H}_{\rm{es}}}$, 
 one obtains a relation of $\epsilon=(p/k')\braket{\bm{S}_{i}\cdot\bm{S}_{j}}$ with $k'=k/N$ where $N$ is the number of Cu2 sites.

\begin{figure}[t]
\includegraphics[scale=0.43]{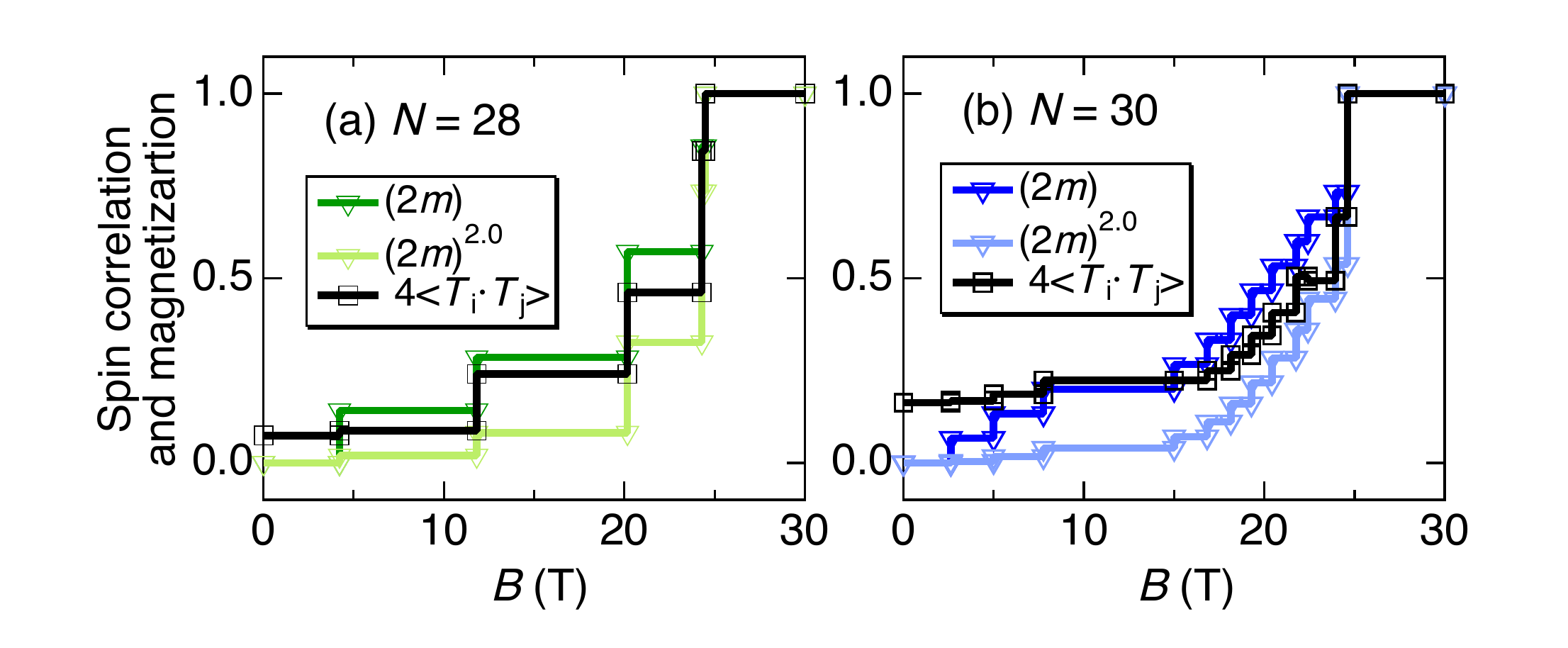}
\caption{Magnetization $m=\braket{T_{i}^{z}}$, $m^{2}$ and the nearest-neighbor spin correlation $\braket{{\bm{T}}_{i}\cdot {\bm{T}}_{j}}$ on the ${\cal J}_1$ bond calculated with the exact diagonalization of the effective model in Fig.\ref{lattice}(c) as a function of the external magnetic field. 
The saturation $m=1/2$ of pseudospins corresponds to the 1/3 magnetization plateau in the original model in Fig.\ref{lattice}(b).
The number of pseudospins is (a) $N=28$ and (b) $30$.
We set ${\cal J}_1=-34.9$ K and ${\cal J}_2=36.5$ K, and also include small further-neighbor couplings \cite{JansonPRL}.
Each curve is normalized so that it becomes unity at the saturation.
\label{lanchoz}}
\end{figure}

The result of the DFT$+U$ calculations is quantitatively analyzed, showing a reasonable agreement with the experimental result in an order of magnitude.
The total energy increase at $\Delta L/L (=\epsilon) = - 1.7\times10^{-3}$ $(- 3.4\times10^{-3})$  is $\Delta E=240$ (539) K per unit cell in the DFT$+U$ calculation.
This sizable increase is dominated by the elastic energy, because the changes in the magnetic exchange energy are of the order of several K as seen in Fig. \ref{DFT}(a).
Considering the increase in the elastic energy per bond as $\Delta E/8=c\epsilon+k'\epsilon^{2}/2$, we obtain the value of the elastic constant $k'$ to be $2.46\times10^{6}$ K per bond, where there are 8 bonds for $J_{1}$ in a unit cell.
The obtained value of $k'$ corresponds to a Young's modulus $\lambda$ of 303 GPa with the relation $\lambda=zk'/V_{\rm{unit}}$ where $z=8$ is the number of $J_{1}$ bonds in a unit cell and $V_{\rm{unit}}=892.125$ \AA$^{3}$ \cite{IshikawaPRL}.
We note that in the DFT+$U$ result, the minimum point of the total energy is shifted slightly from $\epsilon=0$ to $\epsilon=-c/k'=1.25\times 10^{-2}$, which is irrelevant to the magnitude of the magnetostriction. 
The change in the exchange constant $J_{1}$ at $\epsilon = - 3.4\times10^{-3}$ is -5.17 K in the DFT$+U$ calculation.
Considering the linear relation $\Delta J_{1}=-p\epsilon$ as seen in Fig. \ref{lanchoz}, we obtain the value of $p$ to be -1520 K.
Using the obtained values of $k'$ and $p$ and the relation $\epsilon=(p/k')\braket{\bm{S}_{i}\cdot\bm{S}_{j}}$, one obtains a value of $\epsilon_{\rm{DFT}} =- 6.9\times10^{-5}$, where we assumed $\braket{{\bm{S}}_{i}\cdot{\bm{S}}_{j}}=1/9$.
The value of 1/9 for the spin correlation is based on the product of the $S_{\rm{tri}}^z=1/2$ states at the 1/3 magnetization plateau in the spin-1/2 system \cite{JansonPRL}.
The obtained value of $\epsilon_{\rm{DFT}}$ is of the same order of magnitude as 
$\epsilon_{\rm{EXP}} =- 1.7\times10^{-5}$ at the plateau, indicating that the pantograph-like lattice modification is a plausible mechanism.

We finally discuss the observed dependence of $\Delta L/L\propto M^{1.3}$.
First, we compare the present system with other spin systems.
Then, we consider the behavior of the local spin correlations in the present system.
Relation of $\Delta L/L\propto M^{1.0}$ is observed in the dimer-spin systems, SrCu$_{2}$(BO$_{3}$)$_{2}$\cite{NarumiJPSJ, JaimePNAS, RadtkePNAS} and KCuCl$_{3}$ \cite{sawai}, where the pantograph-like lattice change is discussed for Cu dimers.
In these systems, the dominant interaction is the intra-dimer antiferromagnetic coupling, which results in the formation of the singlet ground state below a spin gap \cite{Kageyama}.
Magnetostriction is proportional to the effective number of spin dimers in the excited $S_{\rm{dim}}=1$ state that are transformed from the low-lying $S_{\rm{dim}}=0$ state under magnetic fields, which also is proportional to the magnetization.
In other cases, a dependence in the form of an even function of the magnetization, $\Delta L/L=c_{2}M^{2}+c_{4}M^{4} +\dots$ with $c_{2}$, $c_{4}$, $\dots$ being constants, is usually expected for a variety of magnetic systems based on the symmetry considerations on the magnetic point groups \cite{JaimeNC, Romanov}.

We argue that the local spin correlator $\braket{{\bm{S}}_{i} \cdot {\bm{S}}_{j}}$ on the $J_1$ bond should be responsible as an origin of the present observation of $\Delta L/L\propto M^{1.3}$, 
To calculate it, we resort to the effective model shown in Fig. \ref{lattice}(c) where the low-lying spin-1/2 sector of each trimer is described by the pseudospin-1/2 operator $\bm{T}_{i}$.
Note that $\langle \bm{S}_i\cdot\bm{S}_j\rangle$ on the $J_1$ bond is equal to $(4/9)\langle \bm{T}_i\cdot\bm{T}_j\rangle$ on the ${\cal J}_1$ bond in the effective model.  
Calculated results of $\braket{\bm{T}_{i} \cdot \bm{T}_{j}}$ and $m=\braket{T^{z}_{i}}$ are compared in Figs. \ref{lanchoz}(a) and \ref{lanchoz}(b), which are obtained with exact diagonalization (ED) for finite clusters.
Notice that the curves of the local spin correlation tend to be located in between those of $m^{1.0}$ and $m^{2.0}$ both for $N=28$ and 30 with some exception at low fields where finite-size effects are still found to be large.
This result is consistent with the experimental observation of $\Delta L/L\propto M^{1.3}$ and the relation $\Delta L/L=(p/k')\braket{\bm{S}_{i}\cdot\bm{S}_{j}}$ derived from the exchange striction mechanism.

The observed deviation from the quadratic behavior $\Delta L/L\propto M^2$ can also be interpreted in terms of a spin density wave (SDW) order, whose indications have been observed over a wide range of magnetic field up to 22 T below 3 K in NMR experiment \cite{IshikawaPRL, YoshidaPRB}.
Such an order is also predicted to appear in the effective model in Fig. \ref{lattice}(b) with ${\cal J}_2>0$ (irrespective of the sign of ${\cal J}_1$) \cite{Starykh}.
This order leads to the non-zero covariance $\Delta=\braket{\bm{T}_{i}\cdot\bm{T}_{j}}-m^{2}=\braket{(\bm{T}_{i}-m\hat{z})\cdot(\bm{T}_{j}-m\hat{z})}\neq 0$ as we explain in the following. 
In the effective model, a SDW can be described as $\braket{{T}^{z}_{i}}=m+A\cos(qx/b)$ \cite{Starykh}, where $A$ is the amplitude of the SDW order, $q$ is the wave number with $q=\pi - 2\pi m$, and $x$ is the position of the site $i$ in the $b$ axis of Fig. \ref{lattice}(b).
Note that, on the nearest-neighbor site $j$, $\braket{{T}^{z}_{j}}=m-{\rm sgn}({\cal J}_1)A\cos(qx/b+q/2)$.
Assuming $\braket{(T^{z}_{i}-m)\cdot(T^{z}_{j}-m)}\simeq \braket{T^{z}_{i}-m}\braket{T^{z}_{j}-m}$, one obtains $\Delta \simeq -{\rm sgn}({\cal J}_1)(A^2/2)\cos(q/2)$ after taking the spatial average and neglecting the terms of $\braket{T^{x}_{i}T^{x}_{j}}+\braket{T^{y}_{i}T^{y}_{j}}$.
The experimental data in Figs. \ref{result}(e) and \ref{result}(f) indicate $\Delta>0$, which corresponds to ferromagnetic ${\cal J}_1<0$. 
Even above $T_{\rm{SDW}}$, $\Delta$ is still expected to be non-zero because the short-range correlation develops prior to the long-range one
The present discussion explains the deviation from $\Delta L/L\propto M^2$ both for 4.2 and 2.2 K.

Another possible microscopic origin of $\Delta \neq 0$ is the lateral spin correlation,  $\braket{{T}^{x}_{i}{T}^{x}_{j}}+\braket{{T}^{y}_{i}{T}^{y}_{j}}$, which is neglected in the above discussion.
In the magnon Bose-Einstein condensates (BEC), the lateral spin moment is fixed globally, resulting in the infinite correlation length in the lateral spin correlation which appears in the magnetostriction \cite{zapf2008}.
Even without a long-range order, the pair correlation function is expected to be non-zero at short separations \cite{Hikihara}. 
The spin nematic phase whose indications have been observed in volborthite below 2 K \cite{IshikawaPRL, YoshidaPRB},  is a BEC of bi-magnons.
Though its order parameter, $\braket{T_{i}^{+}T_{j}^{+}}$ \cite{Shannon}, differs from that of the single-magnon BEC, $\braket{T_{i}^{+}}$ \cite{ZapfRMP}, the magnetostriction may still show some anomaly at phase transitions as it is related to the energy on the $J_1$ bonds.

In summary, the single crystalline volborthite is found to show a negative longitudinal magnetostriction  in the $b$ axis up to 45 T with a relation of $\Delta L/L \propto M^{1.3}$ by means of the FBG-based magnetostriction measurement.
The results are discussed in terms of the exchange striction model.
It is argued that the negative magnetostriction arises from the pantograph-like lattice change in the Cu-O-Cu chain in the $b$ axis, strengthening the ferromagnetic exchange coupling in $J_{1}$ bond, which is supported by the DFT+$U$ calculations.
The relation of $\Delta L/L \propto M^{1.3}$ is discussed in terms of the local spin correlator, which is reproduced in the ED of the effective model.
The scope of the future studies includes a possible observation of the signature of the spin nematic phase below 2.0 K and a search for a possible structural transition at higher fields using the high-speed 100 MHz magnetostriction monitor \cite{IkedaFBGHS1, IkedaFBGHS2}.

This work was supported by JSPS KAKENHI Grant-in-Aid for Young Scientists (B) Grant No. 16K17738, Grant-in-Aid for Scientific Research (B) Grant No. 16H04009 and the internal research grant from ISSP, UTokyo. 
OJ was supported by the Austrian Science Fund (FWF) through the Lise Meitner programme, project no. M2050. DFT calculations have been done on the Vienna Scientific Cluster (VSC).

\bibliography{volborthite}

\end{document}